\documentclass[12pt,preprint]{aastex}

\usepackage{lscape}

\begin{document}

\title{Observation of the first gravitational microlensing event in a sparse stellar field : the Tago event}

\author{
A. Fukui\altaffilmark{1},
F. Abe\altaffilmark{1},
K. Ayani\altaffilmark{2}, 
M. Fujii\altaffilmark{4}, 
R. Iizuka\altaffilmark{5}, 
Y. Itow\altaffilmark{1},
K. Kabumoto\altaffilmark{3}, 
K. Kamiya\altaffilmark{1},
T. Kawabata\altaffilmark{1},
S. Kawanomoto\altaffilmark{6},
K. Kinugasa\altaffilmark{7}, 
R. A. Koff\altaffilmark{8},
T. Krajci\altaffilmark{9},
H. Naito\altaffilmark{5},
D. Nogami\altaffilmark{10},
S. Narusawa\altaffilmark{5},
N. Ohishi\altaffilmark{6},
K. Ohnishi\altaffilmark{11},
T. Sumi\altaffilmark{1},
F. Tsumuraya\altaffilmark{5}
}

\altaffiltext{1}{Solar Terrestrial Environment Laboratory, Nagoya University, Nagoya, 464-8601, Japan. 
e-mail:{\tt afukui,abe,itow,kawabata, kkamiya,sumi@stelab.nagoya-u.ac.jp}}
\altaffiltext{2}{Bisei Astronomical Observatory, 1723-70 Ohkura, Bisei, Okayama 714-1411, Japan}
\altaffiltext{3}{Okayama Astronomical Museum, 3037-5 Honjo, Kamogata, Asakuchi, Okayama 719-0232, Japan}
\altaffiltext{4}{ Fujii-Bisei Observatory, 4500 Kurosaki, Tamashima, Okayama 713-8126, Japan}
\altaffiltext{5}{Nishi-Harima Astronomical Observatory, Sayo-cho, Hyogo 679-5313, Japan}
\altaffiltext{6}{National Astronomical Observatory of Japan, 2-21-1 Osawa, Mitaka, Tokyo 181-8588, Japan}
\altaffiltext{7}{ Gunma Astronomical Observatory, 6860-86 Nakayama Takayama, Agatsuma, Gunma 377-0702, Japan}
\altaffiltext{8}{Antelope Hills Observatory, 980 Antelope Drive West Bennett, CO 80102, USA. }
%e-mail:{\tt bob@antelopehillsobservatory.org}
\altaffiltext{9}{Center for Backyard Astrophysics, Cloudcroft, NM 88317, USA.}
%e-mail:{\tt tom\_krajci@tularosa.net}
\altaffiltext{10}{Hida Observatory, Kyoto University, Kamitakara, Gifu 506-1317, Japan. e-mail:{\tt nogami@kwasan.kyoto-u.ac.jp}}
\altaffiltext{11}{Nagano National College of Technology, Tokuma 716, Nagano 381-8550, Japan. e-mail:{\tt kouji.ohnishi@nao.ac.jp}}

%---------------------------------------------

\begin{abstract}

  We report the observation of the first gravitational microlensing
  event in a sparse stellar field, involving the brightest (V=11.4 mag) and closest ($\sim 1 $kpc) source star to date.  This event was discovered by an amateur astronomer,
  A. Tago, on 2006 October 31 as a transient brightening, by $\sim4.5$
  mag during a $\sim$ 15 day period, of a normal A-type star
  (GSC 3656-1328) in the Cassiopeia constellation. 
  Analysis of both spectroscopic
  observations and the light curve indicates that this event was
  caused by gravitational microlensing rather than 
  an intrinsically variable star.  Discovery of this single event over a
  30 year period is roughly consistent with the expected microlensing
  rate for the whole sky down to $V = 12$ mag stars.  However, the
  probability for finding events with such a high magnification ($\sim
  50)$ is much smaller, by a factor $\sim 1/50$, which implies that
  the true event rate may be higher than expected.  This discovery
  indicates the potential of all sky variability surveys, employing
  frequent sampling by telescopes with small apertures and wide fields
  of view, for finding such rare transient events, and using the
  observations to explore galactic disk structure and search for
  exo-planets.
\end{abstract}


%----------------------------------------------

\section{Introduction}

\label{sec:introduction}

The idea that a star's gravity magnifies the light from a perfectly
aligned background source star, so-called gravitational
microlensing, was first presented by \cite{ein36} and developed by
\cite{ref64}.  However, it was thought very difficult to observe such
an event due to its rareness.  \cite{pac86} first estimated realistic
rates for microlensing toward crowded stellar fields, such as the
Large and Small Magellanic Clouds and the Galactic Bulge (GB,
\citealt{pac91}), to study Galactic dark matter, Galactic structure
and extrasolar planets.  The first microlensing candidates were
reported by the MACHO (\citealt{alc93}), EROS (\citealt{aub93}) and OGLE
(\citealt{uda93}) collaborations.  Following this, a few thousands of
microlensing events have been detected to date, mostly towards the GB.
Currently, the OGLE and MOA (\citealt{sum03}) collaborations are
detecting $\sim 600$ events every year.
All of these events have been found only in distant, dense stellar
fields.  This is because the rate of microlensing is very small,
$\sim10^{-5}$ events/star/yr even toward the GB, which is the most
dense stellar field in the sky.  The rate for the whole sky down to
$V=12$ mag for the source stars is estimated to be $\sim 0.05 - 0.2$ events/yr
(\citealt{nem98}, \citealt{han07}).

Here we report the observation of the first microlensing event for a
very close ($\sim 1$ kpc), bright star as a microlensing source that is not within a dense
stellar field.  We present spectroscopic follow-up data, which are used
to judge that this event is most likely due to microlensing rather
than an intrinsically variable star.  We also present observed light
curves to demonstrate that the microlensing hypothesis is plausible.
In \S \ref{sec:discovery} we detail how the event was discovered.  \S
\ref{sec:spectroscopic} contains the spectroscopic analysis, and \S
\ref{sec:photometric} is devoted to the photometric observations and
the light curve analysis.  Discussion and conclusions are given in \S
\ref{sec:conclusion}.

%----------------------------------------

\section{Discovery}

\label{sec:discovery}

On 2006 October 31, an amateur astronomer A. Tago found that the star
GSC 3656-1328 at (R.A., Dec., J2000.0)$=$$(00h09m21.9948s,$
$+54^\circ39'43.832")$, i.e., ($l,b$)=$(116.8158^\circ, -07.7092^\circ)$, near
the Cassiopeia constellation, had brightened by $\sim 4.5$ mag.
He used a commercial digital camera with 70 mm f/3.2 optics, which has
a $\sim$400 $deg^2$ field of view.  This discovery was rapidly
reported to the Central Bureau for Astronomical Telegrams
\footnotemark\footnotetext{\tt http://cfa-www.harvard.edu/iau/cbat.html}
(CBAT; CBET 711) and the Variable Star NETwork
(VSNET;\citealt{kat04}).  Following this, a special alert was
announced by the American Association of Variable Star Observers
\footnotemark\footnotetext{\tt http://www.aavso.org/} (AAVSO; ASN
22).  Meanwhile, Y. Sakurai, an amateur astronomer, independently
discovered the same event.

The source star is a normal A-type star with V=11.4 mag at 1 kpc (CBET
711, 712).  A. Tago had discovered this event during his systematic nova
survey along the Galactic disk, covering $\sim 4500$ $deg^2$ in
total down to $\sim$ 12 mag, and sampling once every other
night.  Such surveys have been conducted for more than 30 years by
many amateur astronomers, including him.  The object was found by
visually comparing successive images with a reference image using the
eye-blinking method.

Spectroscopic and photometric follow-up observations were promptly
obtained, and as the spectra contained no known signatures of variable
stars, the possibility that the brightening was due to microlensing
was reported to the Astronomer's Telegram \footnotemark\footnotetext{\tt
  http://www.astronomerstelegram.org/?read=931} (ATEL; ATEL 931)
web site.

%-----------------------------------------------

\section{Spectroscopic observations}

\label{sec:spectroscopic}

We started our spectroscopic campaign on this variable object 1
night after we received the report on CBET 711.  Five sites were
employed: the Okayama Astrophysical Observatory (OAO), Bisei Astronomical
Observatory (BAO), Fujii Bisei Observatory (FBO), Gunma Astronomical
Observatory (GAO), and Nishi-Harima Astronomical Observatory (NHAO).
For the 2006 November/December period, spectra were obtained as
follows: 2006 November 1 (GAO, FBO, BAO, and OAO), November 4 (FBO and OAO), 
November 8 (FBO and GAO), November 12 (BAO), December 2 (NHAO), and
December 11 (GAO). See Table \ref{tbl:telescope} for instrumental details.

All frames were reduced in the standard manner using IRAF, CCDOPS, and
MaxIm.  The latter two software packages were used for dark
subtraction and flat-field averaging in the case of the FBO data.
Flux calibration using a local standard star was applied to all
spectra except for those obtained at OAO with the HIgh Dispersion
Echelle Spectrograph (HIDES; \citealt{izu99}).  The typical error in
the flux calibration is estimated to be 10\%-20 \%.

\begin{figure}[ht]

\epsscale{0.75}
\includegraphics[angle=-90,scale=0.75,keepaspectratio]{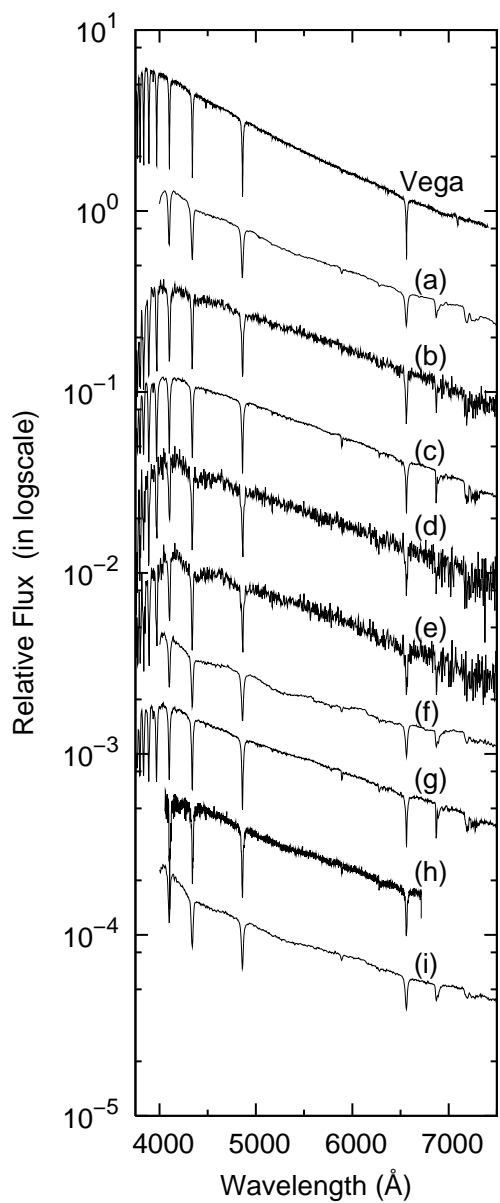}
\caption{ 
Low-dispersion spectra obtained between 2006 November
1, and 2006 December 11, and a spectrum of Vega (A0V).
The vertical axis is the relative flux using a logscale.  The spectra
have been arbitrarily scaled in the vertical direction for visual purposes.
The Vega spectrum was retrieved from the Medium resolution INT
Library of Empirical Spectra (MILES).
(a), (f) and (i) were obtained at GAO. (b), (d) and (e) were taken at FBO.
(c) and (g) were taken at BAO and (h) was at NHAO.  
The blue trend of the continuum did not change within the error of the flux
calibration, and is in good agreement with that of the Vega spectrum.
\label{fig:low-dispersion}}
\end{figure}

Low-dispersion spectra were obtained in the period 2006 November 1 to
2006 December 11.  We continued the campaign for about 1 month after
the object returned to its quiescent state in order to check for any
spectral variation.  All the spectra are displayed in Fig.
\ref{fig:low-dispersion}.  The Vega spectrum shown at the top of this
figure was retrieved from the Medium resolution INT Library of
Empirical Spectra (MILES \footnotemark\footnotetext{\tt
  http://www.ucm.es/info/Astrof/miles/miles.html}).  It is clear that
all these spectra have the same blue trend in the continuum, which is
independent of the brightness of the object.  This trend well matches
that in the Vega spectrum, and the wavelengths at the peak intensities
are also consistently close to that of Vega.  These facts suggest that
the variable object has a spectrum of a normal early A-type dwarf, both
during the bright state and when in quiescence.

\begin{figure}[ht]
\epsscale{0.5}
\includegraphics[angle=-90,scale=0.75,keepaspectratio]{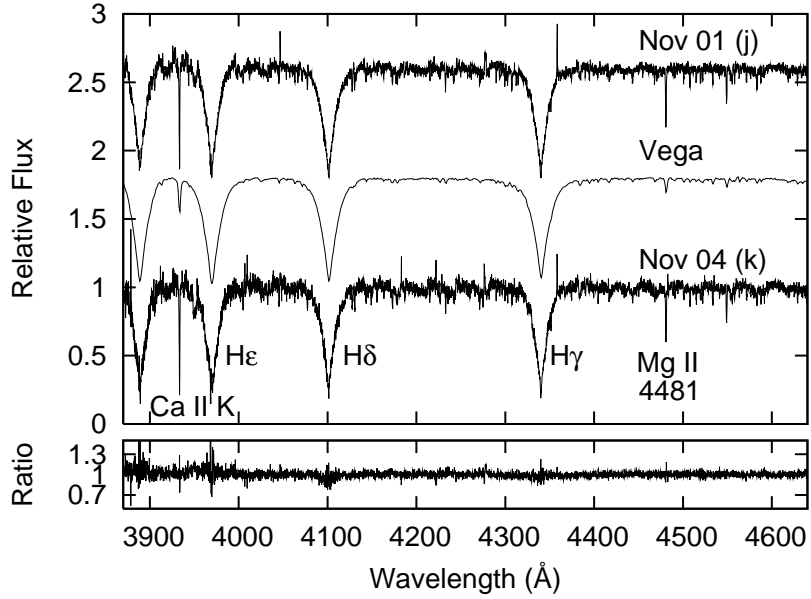}
\caption{ 
High-dispersion spectra of the event star obtained on
2006 November 1 and 4 at the OAO.  We have also included a spectrum
of Vega for comparison purposes: it was obtained at Steward
Observatory using a 90 inch(2.3m) telescope, and is publicly available
on the Web (http://stellar.phys.appstate.edu/stdSO90.html).
These spectra have been normalized to a unity continuum value and
arbitrary shifted in the y axis direction for visual purposes.
The absorption features of H$\gamma, \delta,$ and H$\epsilon$, 
Mg II and Ca II K did not change between the observations and are very
close to those of Vega. The bottom panel shows
the ratio of these two spectra ((j)/(k)).
\label{fig:high-dispersion}}
\end{figure}

The high-dispersion spectra normalized to a unity continuum value are
shown in Fig. \ref{fig:high-dispersion}.  These two spectra were
obtained on 2006 November 1 and 4, when the object was about 6.4 and 1.9
times brighter in V compared with quiescence, respectively (see Fig.  \ref{fig:lightcurve}).
While the object's brightnesses were
different, both spectra contain the same absorption-line features
(Balmer, Mg II and Ca II K) in terms of depth, width, equivalent
width, and so on.  This is clearly demonstrated by the ratio of these
two spectra, displayed in the bottom panel of Fig.
\ref{fig:high-dispersion}.  In addition, the absorption-line features
are quite close to those in the Vega spectrum, and the differences of
the widths and depths of these lines between the target and Vega
should be attributable to the difference between the projected rotation
velocities of the two objects.  Note that some of the end effects
resulting from the diffraction orders could not be completely
eliminated during the analysis and have remained in the continuum.

In summary, our observations revealed that the optical spectrum of the
variable object did not vary from when the object was $\sim$6.4 times
brighter than its quiescent state until it returned to this state, and
that the spectrum is almost the same as that of Vega.  Apart from
microlensing, we are not aware of another large-amplitude brightening
phenomenon with a timescale of days that does not change the object's
physical properties, such as the temperature and density.  For
example, stellar flares and nova explosions are accompanied by strong
enhancement of Balmer emission lines, and dwarf nova-type outbursts
show a decay of the Balmer emission lines.  The invariance of the
variable's spectral features, and the fact that they are close to
Vega's spectrum, provide firm evidence that the source object is a
normal main-sequence star of late-B to early-A type, and that this
brightening event was caused by gravitational microlensing.

%----------------------------------------

\section{Photometric observations and light curve analysis}

\label{sec:photometric}

Photometric observations were made by various observers before and
after the alert, using both commercial digital cameras and 
scientific CCDs.  A. Tago, Y. Sakurai, and Y. Sugawara obtained 22, 9,
and 1 JPEG images, respectively, using digital cameras (digicam data).
The observation times of data posted by A. Tago on the CBAT were initially
found to be wrong, but were then corrected (A. Tago 2007, private
communication).  R.A.K and T.K. both obtained CCD
images of the variable and posted them on the AAVSO website.
The former provided 3635 V-band frames, while the latter took
3472 frames without a filter.  I-band (13 CCD images) and V-band (24
CCD images) were taken by the north site (Hawaii) of ASAS
(\citealt{poj02}) by chance as part their survey of the north sky, and
not prompted by the alert.

The scientific CCD images, in the FITS format, were reduced in the
standard manner, and photometry of the variable was extracted.
However, the JPEG images taken by commercial digital cameras were
reduced with special care because both dark and flat-field images were
not available, and the flux linearity was affected by the lossy
compression method used.  We estimated the uncertainties due to these
effects from the scatter of a standard star as follows.

First, using three colors we added all the flux in the three images to
produce a single JPEG image and then converted this into a FITS
formatted image.  Then about 100 stars in a $5^\circ \times 7^\circ$
area around the event star in each image were cross-referenced with
stars in the Tycho-2 catalog (\citealt{hog00}) for comparison purposes.
These stars were also chosen to have a similar color, B-V = $0.10 \pm
0.1$, and they ranged from 6.9 to 12 mag in $V_T$, defined by Tycho-2.
The gain uniformity of the digital cameras in this region was
determined by examining sky-flat images taken by the cameras, and it
was found that they were flat over the area to better than $3\%$.  We
then performed aperture photometry on these stars using IRAF, and
determined the relation between the instrumental magnitudes and the $V_T$
magnitudes by fitting a quadratic function, taking errors in both
magnitudes into account.  Since the IRAF errors are relative values,
we renormalized these errors by a factor so that $\chi^2 / dof$ was
unity.  The IRAF error of the event star was also renormalized by the
same factor.  Finally, the IRAF magnitudes and associated errors of the
event star were converted into $V_T$ magnitudes using these functions
with their errors.

To test whether the photometric light curve was consistent with the
microlensing model or not, we fitted the combined light curve with a
simple point-lens and point-source model, characterized by the three
parameters $u_{0}$, $t_{0}$ and $t_{E}$ (\citealt{pac86}), where
$u_{0}$ is the minimum impact parameter corresponding to the
source-lens angular separation in units of the angular Einstein radius $\theta_{E}$, $t_{0}$ is the
time of maximum magnification, and $t_{E}$ is the event timescale, defined as
$\theta_{E} / \mu$ where $\mu$ is the relative proper motion between the source and lens stars.
The baseline flux of the source star and any blended flux in each data
set were also allowed to be free parameters in the fitting procedure.

\begin{figure}[ht]
\epsscale{0.5}
\includegraphics[angle=-90,scale=0.6,keepaspectratio]{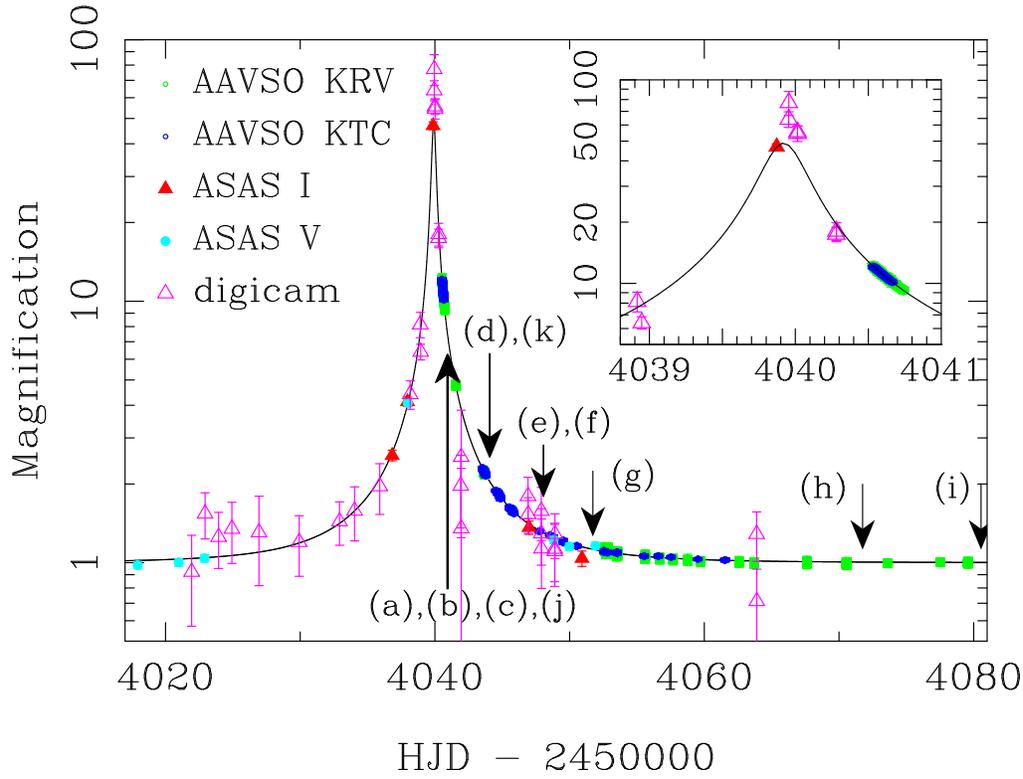}
\caption{
The light curve of the Tago event. The solid line indicates the best fit 
point-lens and point-source model.  Errors in the AAVSO data
have been re-normalized as noted in the text. The arrows (a) - (k)
indicate the epochs of the spectroscopic measurements depicted in 
Fig \ref{fig:low-dispersion} and Fig \ref{fig:high-dispersion}.
\label{fig:lightcurve}}
\end{figure}

In Table \ref{tbl:param} we list the resultant parameters and the errors
that were obtained for the different
data subsets by checking the systematics within each one.  The
uncertainties in the ASAS and AAVSO data were estimated to be equal to
the rms of their baseline values.  These were determined
using only those observations separated by an interval of at least
$3 t_{E}$ from the peak magnification, where the amplitudes are
expected to be sufficiently low.  The value for $t_{E}$ was determined
from the fitting parameters of the digicam data set.  The $\chi^2/dof$ in
each data set is close to unity except for the AAVSO data.  We found
that the AAVSO points were scattered about the fitted curve
significantly more than expected from the estimated errors, with a timescale of $<$ 1 day.
This is because the reduction and calibration of
the data sets were not optimal (e.g. no air mass corrections were
made), and the errors were also underestimated for some reason.
Other than this, one can see that the microlensing model provides a
satisfactory fit to all the data sets  except for some digicam data points near the peak 
(refer to Fig.\ref{fig:lightcurve}). We suspect this deviation is due to the systematic errors
from lossy compression, and we think this has an insignificant effect on the conclusion of this paper.
Consequently, we renormalized the errors of
the R.A.K and T.K. data sets so that $\chi^2/dof$ in each one was unity,
where the fitted model values were determined using all data sets,
including the unnormalized AAVSO data (see "*" in Table
\ref{tbl:param}).  The blended flux parameter for each data subset is
nearly negligible in comparison with the source flux, except for the
ASAS data set.  In this case there were contributions from an
unresolved neighboring star due to the low resolution.

Fig. \ref{fig:lightcurve} shows the complete photometric light curve,
including the renormalized AAVSO data, and the best fit model to all
the data.  As shown in Table \ref{tbl:param}, the differences in
parameters for the various data subsets are relatively small, e.g.,
$\sim 3$ days in $t_{\rm E}$ at most, which is roughly the size of the
statistical errors.  Therefore we conclude that systematic biases in
all data sets are negligible, and the aggregate light curve can be
represented well by a point-lens and point-source microlensing model.

%-----------------------------------------

\section{Discussion \& Conclusion}

\label{sec:conclusion}

We have reported the observation of the first microlensing event in a
sparse stellar field, involving the brightest (V=11.4 mag) and closest
(1 kpc) source star to date.  The spectra of the source star and the
overall light curve show that this event cannot have been caused by any known
mechanism for intrinsically variable stars, but it can be naturally
explained as a microlensing event.

\cite{gau07} also reported the discovery of the same event, and also
concluded that this event is most likely due to microlensing.  A subset
of the data used in this paper is in common with theirs, but our
analysis was carried out independently, and we have original data
taken by amateur astronomers, which includes coverage of the peak
magnification region.

The rate for the microlensing event described here is estimated to be
very small at $\sim 0.05 - 0.2$ events/yr over the whole sky
(\citealt{nem98}, \citealt{han07}), so it seems reasonable for one such event to happen
in a 30 year period.
However, this event is a high amplification one, so the corresponding
rate is much smaller ($\sim 1/50$) than that given above.  This may be
just good luck, or the true event rate may be higher than expected but low-
amplitude events may be missed by the current nova surveys with digital
cameras, where the detection efficiency is uncertain.  Such a
possibility can be addressed by systematic all-sky surveys featuring
frequent sampling with small apertures and wide fields of view, such
as LOTIS (\citealt{par02}), ROTSE (\citealt{woz04}) and ASAS
(\citealt{poj02}).  We can expect to detect some microlensing
events/yr for all-sky surveys down to $V = 15$ mag (\citealt{nem98}, \citealt{han07}).

Searching for exo-planets (\citealt{mao91}) in very bright microlensing events 
is also a worthwhile exercise. 
Although the expected number of these events is small, e.g. 1 planet every 4 years if assuming 5 microlensing events observed each year by an all-sky survey and 5\% planet abundance,
the detection efficiency for planets in such bright events is much higher than for faint events,
because the photometric precision of a bright event is higher (Peale 2001). 
In addition, such high precision can be achieved even by small-aperture telescopes distributed around the world.
Even if we can detect only one planet in such an all-sky survey in 2-10 years, it will be the first
planet discovered by microlensing in the solar neighborhood which can
be confirmed by other methods  such as spectroscopy or direct imaging after the peak amplification phase has ended.
This will be very important to convince people that the planetary signal in
the microlensing that we are also seeing towards the bulge is actually caused by the planet.

This discovery shows the potential of all-sky variability surveys
using cheap, small telescopes for finding rare transient events, which
in turn will provide information on disk structure and possible
exoplanets.

%---------------------------------------

\acknowledgments

We would like to express our deep gratitude to A. Tago, Y. Sakurai,
and Y. Sugawara for providing the significant data we used for this
analysis, and acknowledge T. Fukashima, T. Izumi, K. Kadota, S.
Kaneko, S. Kiyota, K. Nakajima, H. Nishimura, H. Maehara, Y. Okamoto,
S. Tago, M. Yamamoto, J. Yokomichi, and many other observers for
providing their observations or valuable information.  And, special
thanks to H. Yamaoka for providing support for the data collection.  We also
would like to thank the many amateur observers who obtained valuable data and
made them publicly available by sending them to VSNET, AAVSO, and VSOLJ.
We acknowledge G. Pojmanski and D. Szczygiel for providing the data of
ASAS.  This work was supported by Grant-in-Aids (15540240, 17340074,
18253002, 18749004, and 18740119) from MEXT and JSPS.

%---------------------------------------------------

\begin{deluxetable}{rrlr}
\tabletypesize{\scriptsize}
\tablecaption{Telescopes and instruments used for spectroscopic observations.
\label{tbl:telescope}}
\tablewidth{0pt}
\tablehead{
Site$^{a}$ &  Tel. & Instruments   &  WR$^{b}$ (A) \\
}

\startdata
OAO &  188cm & HIDES$^{c}$                  & 0.07 \\
BAO &  101cm & Low Dispersion Spectrograph  &   5 \\
FBO &   28cm & Low Dispersion Spectrograph  &  10 \\
GAO &  150cm & Low Dispersion Spectrograph  & 13-14 \\
NHAO&  200cm & MALLS$^{d}$                  &   6 \\
\enddata

\tablecomments{
$^{a}$ OAO: Okayama Astrophysical Observatory,
BAO: Bisei Astronomical Observatory,
FBO: Fujii Bisei Observatory,
GAO: Gunma Astronomical Observatory,
NHAO: Nishi-Harima Astronomical Observaotry,
$^{b}$ WR: Wavelength resolution,
$^{c}$ HIDES: HIgh Dispersion Echelle Spectrograph (Izumiura 1999),
$^{d}$ MALLS: Medium And Low-dispersion Long slit Spectrograph
}

\end{deluxetable}

\begin{deluxetable}{lrrrrrr}
\tabletypesize{\scriptsize}
\tablecaption{Lensing fit parameters.
\label{tbl:param}}
\tablewidth{0pt}
\tablehead{
dataset & $t_0 (HJD - 2450000)$ & $u_{\rm 0}$ &  $t_{\rm E} (day)$ &   $N$ & $\chi^2/dof$ \\
}

\startdata
digicam               & 4039.93 (0.02)  &  0.0197 (0.0134) & 4.70 (2.87)  & 32   &  1.04\\
ASAS               & 4039.99 (0.13)  &  0.0221 (0.0129) &  6.20  (0.95)  & 37   &  1.08\\
AAVSO              & 4039.87 (0.006) &  0.0184 (0.0029) & 6.98 (0.02)  & 7107 &  6.75\\
AAVSO$^*$          & 4039.98 (0.02)  &  0.0367 (0.0038) &  7.74  (0.10)  & 7107 &  0.97\\
ASAS+AAVSO$^*$           & 4039.92 (0.005)  &  0.0229 (0.0014) &  7.53  (0.05)  & 7144   &  0.98\\
digicam+AAVSO$^*$ & 4039.89 (0.004) & 0.0107 (0.0018) & 7.48 (0.004) & 7139 & 0.98\\
digicam+ASAS & 4039.98 (0.01) & 0.0148 (0.0020) & 7.24 (0.66) & 69 & 1.27\\
digicam+ASAS+AAVSO & 4039.88 (0.002) & 0.0199 (0.0010) & 6.99 (0.02) & 7176 & 6.71\\
digicam+ASAS+AAVSO$^*$     & 4039.91 (0.003)  &  0.0189 (0.0011) &  7.53  (0.04)  & 7176 & 0.99\\
\enddata

\tablecomments{
digicam: Data from digital cameras. ASAS: ASAS I and V band. AAVSO: CCD data from R.A.K. and T.K.
*: Errors in R.A.K. and T.K. data sets  are rescaled independently by multiplying the factors so that $\chi^2/dof$ in each data set was unity, calculated from a model fitted by using all data sets including the unnormalized AAVSO data. The numbers in parentheses represent 1-$\sigma$ statistical
errors which corresponding to the values where $\Delta \chi^{2}$ is 1.
}

\end{deluxetable}

%--------------------------------------------

\begin{thebibliography}{}
\bibitem[Alcock et al.(1993)]{alc93}Alcock, C. et al. 1993 Nature 365, 621
\bibitem[Aubourg et al.(1999)]{aub93}Aubourg, E., et al. 1993, Nature, 365, 623
%\bibitem[Bond et al.(2004)]{bon04}Bond, I, et al. 2004, \apj, 606, L155
\bibitem[Einstein (1936)]{ein36} A. Einstein, 1936, Science, 84, 506
\bibitem[Gaudi et al.(2007)]{gau07}Gaudi, S. et al. 2007, \apj, submitted (astro-ph/0703125v2)
\bibitem[Han (2007)]{han07}Han, C. 2007, preprint (astro-ph/0708.1215v1)
\bibitem[H{\o}g et al.(2000)]{hog00}H{\o}g E. et al. 2000, A\&A, 355, L27
\bibitem[Izumiura (1999)]{izu99}Izumiura, H., 1999, in Proc. 4th East Asian Meeting on Astronomy, ed. P.S. Chen (Kunming: Yunnan Observatory), 77
\bibitem[Kato et al. (2004)]{kat04}Kato, T., Uemura, M., Ishioka, R., Nagami, D., Kunjaya, C., Baba, H., Yamaoka, H., 2004, PASJ, 56, S1
\bibitem[Mao \& Paczy\'{n}ski(1991)]{mao91}Mao, S. \& Paczy\'{n}ski, B. 1991, \apj, 374, L37
\bibitem[Marshall (1997)]{nem97}Marshall, S., 1997, BAAS, 191 (48.15)
\bibitem[Nemiroff (1998)]{nem98}Nemiroff, R.J., 1998, \apj, 509, 39
\bibitem[Paczy\'{n}ski(1986)]{pac86}Paczy\'{n}ski, B. 1986, \apj, 304,1
\bibitem[Paczy\'{n}ski(1991)]{pac91}Paczy\'{n}ski, B. 1991, \apj, 371, L63
\bibitem[Park et al. (2002)]{par02}Park, H. S. et al. 2002, \apj, 571, L131
\bibitem[Peale (2001)]{pea01}Peale, S. J. 2001, \apj, 552, 889
\bibitem[Pojma\'{n}ski (2002)]{poj02}Pojma\'{n}ski, 2002, Acta Astron., 52, 397
\bibitem[Refsdal (1964)]{ref64} Refsdal, S. 1964, MNRAS, 128, 295
\bibitem[Udalski et al.(1993)]{uda93}Udalski et al., 1993, Acta Astronomica, 43,  69
\bibitem[Sumi et al.(2003)]{sum03}Sumi, T. et al. 2003, ApJ, 591, 204
\bibitem[Wo\'{z}niak et al.(2004)]{woz04}Wo\'{z}niak, P. R. et al. 2004, \aj, 127, 2436
\end{thebibliography}
\end{document}